
\input harvmac
\input diagrams.sty
\noblackbox
\newcount\figno
\figno=0
\def\fig#1#2#3{
\par\begingroup\parindent=0pt\leftskip=1cm\rightskip=1cm\parindent=0pt
\baselineskip=11pt
\global\advance\figno by 1
\midinsert
\epsfxsize=#3
\centerline{\epsfbox{#2}}
\vskip -3cm
\vskip 12pt
\endinsert\endgroup\par}
\def\figlabel#1{\xdef#1{\the\figno}}
\def\pano{\par\noindent}
\def\smno{\smallskip\noindent}
\def\meno{\medskip\noindent}
\def\bigno{\bigskip\noindent}
\font\cmss=cmss10
\font\cmsss=cmss10 at 7pt
\def\rlx{\relax\leavevmode}
\def\inbar{\vrule height1.5ex width.4pt depth0pt}
\def\IC{\relax\,\hbox{$\inbar\kern-.3em{\rm C}$}}
\def\IR{\relax{\rm I\kern-.18em R}}
\def\IN{\relax{\rm I\kern-.18em N}}
\def\IP{\relax{\rm I\kern-.18em P}}
\def\ZZ{\rlx\leavevmode\ifmmode\mathchoice{\hbox{\cmss Z\kern-.4em Z}}
 {\hbox{\cmss Z\kern-.4em Z}}{\lower.9pt\hbox{\cmsss Z\kern-.36em Z}}
 {\lower1.2pt\hbox{\cmsss Z\kern-.36em Z}}\else{\cmss Z\kern-.4em Z}\fi}
\def\narrowplus{\kern -.04truein + \kern -.03truein}
\def\narrowminus{- \kern -.04truein}
\def\narrowminussub{\kern -.02truein - \kern -.01truein}
\def\a{\alpha}

\def\o#1{\overline{#1}}

\def\la{\langle}
\def\ra{\rangle}

\def\cN{${\cal N}$}

\def\typeo{type$\ {\rm 0B}$}

\def\typeb{type$\ {\rm IIB}$}

\def\rN{\rm N}

\def\frac#1#2{{#1\over #2}}
\def\sqr#1#2{{\vcenter{\vbox{\hrule height.#2pt
 \hbox{\vrule width.#2pt height#1pt \kern#1pt
 \vrule width.#2pt}\hrule height.#2pt}}}}
\def\square
 {\mathop{\mathchoice{\sqr{12}{15}}{\sqr{9}{12}}{\sqr{6.3}{9}}{\sqr{4.5}{9}}}}


\def\drawbox#1#2{\hrule height#2pt 
        \hbox{\vrule width#2pt height#1pt \kern#1pt 
              \vrule width#2pt}
              \hrule height#2pt}

\def\Fund#1#2{\vcenter{\vbox{\drawbox{#1}{#2}}}}
\def\Asym#1#2{\vcenter{\vbox{\drawbox{#1}{#2}
              \kern-#2pt       
              \drawbox{#1}{#2}}}}

\def\funda{\Fund{6.5}{0.4}}
\def\asymm{\Asym{6.5}{0.4}}

\def\bfunda{\overline{\funda}}
\def\basymm{\overline{\asymm}}


\def\zwei#1#2{ \left(\matrix{  #1 \cr #2 \cr  }\right) }
\def\drei#1#2#3{ \left( #1 \matrix{ #2 \cr  #3 \cr }\right) }
\def\funf#1#2#3#4#5{ \left( #1 \matrix{ #2 & #3\cr  #4 & #5 \cr }\right) }
\def\vier#1#2#3#4{ \left(\matrix{  #1 &  #2 \cr #3 & #4 \cr }\right) } 


\lref\rkarch{A. Karch, D. L\"ust and D. Smith,
{\it Equivalence of Geometric Engineering and Hanany-Witten via Fractional
Branes}, Nucl.\ Phys.\ {\bf B533} (1998) 348, hep-th/9803232.}

\lref\rhanur{A. Hanany and A.M. Uranga,
{\it Brane Boxes and Branes on Singularities},
hep-th/9805139.}

\lref\randreas{B. Andreas, G. Curio and D. L\"ust,
{\it The Neveu-Schwarz Five-brane and its Dual Geometries},
JHEP {\bf 9810} (1998) 022, hep-th/9807008.}

\lref\rdasmuk{K. Dasgupta and S. Mukhi,
{\it Brane Constructions, Conifolds and M-theory},
hep-th/9811139.}

\lref\ragana{M. Aganagic, A. Karch, D. L\"ust and A. Miemiec,
{\it Mirror Symmetries for Brane Configurations and Branes at Singularities},
hep-th/9903093.}

\lref\rhanzaf{A. Hanany and A. Zaffaroni,
{\it On the Realization of Chiral Four-dimensional Gauge Theories Using Branes},
hep-th/9801134.}

\lref\rhanwit{A. Hanany and E. Witten,
{\it Type IIB Superstrings, BPS Monopoles and Three-dimensional Gauge Dynamics},
Nucl.\ Phys.\ {\bf B492} (1997) 152, hep-th/9611230.}

\lref\rdougmoor{M.R. Douglas and G. Moore,
{\it D-branes, Quivers and ALE Instantons},
hep-th/9603167.}

\lref\rbersh{M. Bershadsky, C. Vafa and V. Sadov,
{\it D-strings on D Manifolds},
Nucl.\ Phys.\ {\bf B463} (1996) 398, hep-th/9510225.}

\lref\relitzur{S. Elitzur, A. Giveon and D. Kutasov,
{\it Branes and $N=1$ Duality in String Theory},
Phys.\ Lett.\ {\bf B400} (1997) 269, hep-th/9702014.}

\lref\rkeha{A. Kehagias,
{\it New Type IIB Vacua and their F Theory Interpretation},
Phys.\ Lett.\ {\bf B435} (1998) 337, hep-th/9805131.}

\lref\rmalda{J. M. Maldacena, {\it The Large N Limit of Superconformal 
Field Theories and Supergravity}, Adv.Theor.Math.Phys. 2 (1998) 231,
hep-th/9711200.}
 
\lref\rnekr{N. Nekrasov and S.L. Shatashvili, {\it On Non-Supersymmetric
    CFT in Four Dimensions}, hep-th/9902110.}

\lref\rpoly{A.M. Polyakov, {\it The Wall of the Cave}, 
Int.J.Mod.Phys. {\bf A14} (1999) 645, hep-th/9809057.}

\lref\rframp{P.H. Frampton and C. Vafa, {\it Conformal Approach to Particle
   Phenomenology}, hep-th/9903226 \semi
    P.H. Frampton, {\it Conformality and Gauge Coupling Unification},
    hep-th/9905042}

\lref\rcsaki{C. Cs\' aki, W. Skiba and J. Terning, {\it $\beta$ Functions
of Orbifold Theories and the Hierarchy Problem}, hep-th/9906057.}

\lref\rbkv{M. Bershadsky, Z. Kakushadze and C. Vafa, {\it String Expansion
as Large N Expansion of Gauge Theories}, Nucl.Phys. {\bf B523} (1998) 59,
hep-th/9803076 \semi
M. Bershadsky and A. Johansen, {\it Large N Limit of Orbifold Field Theories}
Nucl.\ Phys.\ {\bf B536} (1998) 141, hep-th/9803249 .}

\lref\rkakor{Z. Kakushadze, {\it Gauge Theories from Orientifolds and 
Large N Limit}, Nucl. Phys. {\bf B529} (1998) 157, hep-th/9803214\semi
{\it On Large N Gauge Theories from Orientifolds},
Phys. Rev. {\bf D58} (1998) 106003, hep-th/9804184 \semi
{\it Anomaly Free Non-Supersymmetric Large N Gauge Theories from 
Orientifolds}, Phys. Rev. {\bf D59} (1999) 045007, hep-th/9806091\semi
{\it Large N Gauge Theories from Orientifolds with NS-NS B-flux}, 
Nucl. Phys. {\bf B544} (1999) 265, hep-th/9808048.}

\lref\rlnv{A. Lawrence, N. Nekrasov and C. Vafa, {\it On Conformal Field
Theories in Four Dimensions}, Nucl.\ Phys.\ {\bf B533} (1998) 199,
hep-th/9803076.}

\lref\rsilver{S. Kachru and E. Silverstein, {\it 4d Conformal Field Theories
and Strings on Orbifolds}, Phys. Rev. Lett. 80 (1998) 4855, hep-th/9802183.}

\lref\rkol{A. Armoni and B. Kol, {\it Non-Supersymmetric Large N Gauge
Theories from Type 0 Brane Configurations}, hep-th/9906081.}

\lref\rschmaltz{M. Schmaltz, {\it Duality of Non-Supersymmetric Large N Gauge
Theories}, Phys. Rev. {\bf D59} (1999) 105018, hep-th/9805218.}

\lref\rhanany{A. Hanany and E. Witten, {\it TypeIIB Superstrings, BPS 
monopoles and Three Dimensional Gauge Dynamics}, 
Nucl.\ Phys.\ {\bf B492} (1997) 152, hep-th/9611230.}

\lref\rwitten{E. Witten, {\it Solutions of Four-Dimensional Field Theories
Via M-Theory}, Nucl.\ Phys.\ {\bf B500} (1997) 3, hep-th/9703166.}

\lref\rparur{J. Park and A.M. Uranga, {\it A Note on Superconformal N=2
           theories and Orientifolds}, Nucl.Phys. {\bf B542} (1999) 139,
         hep-th/9808161.}

\lref\rprasad{
M.K. Prasad, {\it Equivalence of Eguchi-Hanson metric to two 
center Gibbons-Hawking metric}, Phys.\ Lett.\  {\bf B83} (1979) 310.}

\lref\riban{L.E. Ib\'a\~nez, R. Rabadan and A.M. Uranga, {\it Anomalous U(1)s in
Type I and TypeIIB D=4, N=1 string vacua}, Nucl.Phys. {\bf B542} (1999) 112,
hep-th/9808139.}

\lref\rbillo{M. Billo, B. Craps and F. Roose, {\it On D-branes in
Type 0 String Theory}, hep-th/9902196.}

\lref\rbfl{R. Blumenhagen, A. Font and D. L\"ust, {\it Tachyon-free
Orientifolds of Type 0B Strings in Various Dimensions}, hep-th/9904069.}

\lref\rkletsya{I.R. Klebanov and A.A. Tseytlin, {\it D-Branes and Dual Gauge 
theories in Type 0 Strings}, Nucl.Phys. {\bf B546} (1999) 155, 
hep-th/9811035.} 

\lref\rkletsyb{I.R. Klebanov and A.A. Tseytlin, {\it A Non-supersymmetric
Large N CFT from Type 0 String Theory}, JHEP {\bf 9903} (1999) 015,
hep-th/9901101.} 

\lref\rkletsyc{I.R. Klebanov and A.A. Tseytlin, {\it Asymptotic Freedom
and Infrared Behavior in the Type 0 String Approach to
Gauge Theory},  Nucl.\ Phys.\ {\bf B547} (1999) 143, hep-th/9812089}

\lref\rminahan{J.A. Minahan, {\it Asymptotic Freedom and Confinement from
Type 0 String Theory}, JHEP {\bf 9904} (1999) 007, hep-th/9902074}

\lref\rgarousi{M.R. Garousi, {\it String Scattering from D-Branes in Type 0
Theories}, hep-th/9901085.}

\lref\rzarembo{K. Zarembo, {\it Coleman-Weinberg Mechanism and Interaction
of D3-Branes in Type 0 String Theory}, hep-th/9901106.} 

\lref\rtsz{A.A. Tseytlin and K. Zarembo, {\it Effective Potentials in Non-
Supersymmetric $SU(N)\times SU(N)$ Gauge Theory and Interactions of Type 0
D3-Branes}, hep-th/9902095.}

\lref\rcosta{M.S. Costa, {\it Intersecting D-Branes and Black Holes in 
Type 0 String Theory}, JHEP {\bf 9904} (1999) 016, hep-th/9903128.}

\lref\rleigh{R.G. Leigh and M. Rozali, {\it Brane Boxes, Anomalies, 
Bending and Tadpoles}, Phys.Rev. {\bf D59} (1999) 026004, hep-th/9807082.}

\lref\rmv{M.E. Machacek and M.T. Vaughn, {\it Two-Loop Renormalization 
Group Equation in a General Quantum Field Theory I: Wave Function
Renormalization}, Nucl.Phys. {\bf B222} (1983) 83.}

\lref\rbergab{O. Bergman and M.R. Gaberdiel, {\it A Non-Supersymmetric Open 
String Theory and S-Duality}, Nucl.Phys. {\bf B499} (1997) 183, 
hep-th/9701137.}

\lref\rbergman{O. Bergman and M.R. Gaberdiel, {\it Dualities of Type 0
Strings}, hep-th/9906055.}

\lref\rangel{C. Angelantonj, {\it Non-Tachyonic Open Descendants of the 
0B String Theory}, Phys.Lett. {\bf B444} (1998) 309, hep-th/9810214.}

\lref\rsagbi{A. Sagnotti, M. Bianchi,
{\it On the Systematics of Open String Theories},
Phys.\ Lett.\ {\bf B247} (1990) 517}

\lref\rsagn{A. Sagnotti, {\it Some Properties of Open String Theories},
 hep-th/95090808 \semi
A. Sagnotti, {\it Surprises in Open String Perturbation Theory},
hep-th/9702093.}

\lref\rkachru{S. Kachru, J. Kumar and E. Silverstein, {\it Vacuum Energy 
  Cancellation in a Nonsupersymmetric Strings}, 
    Phys.Rev. {\bf D59} (1999) 106004, hep-th/9807076. }

\lref\rklewit{I.R. Klebanov and E. Witten, {\it Superconformal field 
theory on three-branes at a Calabi-Yau singularity}, 
Nucl. Phys. {\bf B536} (1998) 199, hep-th/9807080.}

\lref\ruranga{A.M. Uranga, {\it Brane configurations for branes at 
conifolds}, JHEP {\bf 9901} (1999) 022, hep-th/9811004.}

\lref\rfisus{W.\ Fischler and L.\ Susskind, 
{\it Dilaton Tadpoles, String Condensates, and Scale Invariance}, 
Phys.\ Lett.\ {\bf B171} (1986) 383;  
{\it Dilaton Tadpoles, String Condensates, and Scale Invariance II}, 
Phys.\ Lett.\ {\bf B173} (1986) 262.}  

\lref\rseibwit{L. Dixon and J. Harvey, {\it String Theories in Ten Dimensions
Without Space-Time Supersymmetry}, Nucl.\ Phys.\ {\bf B274} (1986) 93 \semi
N. Seiberg and E. Witten, {\it Spin Structures in String Theory}, 
Nucl.\ Phys.\ {\bf B276} (1986) 272. }

\lref\roz{ M. Alishahiha, A. Brandhuber and Y. Oz, 
{\it Branes at Singularities in Type 0 String Theory}, 
     JHEP {\bf 9905} (1999) 024, hep-th/9903186.}

\lref\rtatar{K. Oh and R. Tatar, {\it Branes at Orbifolded Conifold 
Singularities and Supersymmetric Gauge Field Theories}, hep-th/9906012.}

\Title{\vbox{\hbox{hep--th/9906101}
 \hbox{HUB--EP--99/28}}}
{\vbox{ \vskip -2cm
       \hbox{Non-Supersymmetric Gauge Theories from}  
\vskip 0.4cm
               \hbox{\phantom{Niee}D-Branes in Type 0 String Theory}}}
\centerline{Ralph Blumenhagen${}^1$, Anamar\'{\i}a Font${}^2$ and 
Dieter L\"ust${}^3$}
\bigskip
\centerline{\it ${}^{1,3}$ Humboldt-Universit\"at Berlin, Institut f\"ur 
Physik,}
\centerline{\it  Invalidenstrasse 110, 10115 Berlin, Germany }
\smallskip
\centerline{\it ${}^2$ Departamento de F\'{\i}sica, Facultad de Ciencias,
 Universidad Central de Venezuela,}
\centerline{\it A.P. 20513, Caracas 1020-A, Venezuela }
\centerline{\it and}
\centerline{\it Centro de Astrof\'{\i}sica Te\'orica, Facultad de Ciencias,}
\centerline{\it Universidad de Los Andes, Venezuela} 
\smallskip
\bigskip
\bigskip
\centerline{\bf Abstract}
\noindent
We construct non-supersymmetric four dimensional gauge theories arising 
as effective theories of D-branes placed on various singularities in Type 0B 
string theory. We mostly
focus on models which are conformal in the large N limit and present
both examples appearing on self-dual D3-branes on orbifold singularities
and examples including orientifold planes. Moreover, we derive 
type 0 Hanany-Witten
setups 
with NS 5-branes
intersected by D-branes
and the corresponding rules for determining the massless spectra.
Finally, we discuss possible duality symmetries 
(Seiberg-duality) for non-supersymmetric
gauge theories within this framework.

\footnote{}
{\pano
${}^1$ e--mail:\ blumenha@physik.hu-berlin.de
\pano
${}^1$ e--mail:\ afont@fisica.ciens.ucv.ve
\pano
${}^3$ e--mail:\ luest@physik.hu-berlin.de
\pano}
\Date{06/99}


\newsec{Introduction}

In recent years we have 
gained new insights about non-perturbative
aspects of supersymmetric gauge theories by studying them as effective
low energy theories arising on D-branes in string theory.
This includes for instance the derivation of the \cN=2 Seiberg-Witten 
curve
from the embedding of five-branes in M-theory \rwitten, as well as, the 
Maldacena conjecture \rmalda\ stating that conformal gauge theories arising
on parallel D3-branes in \typeb\ string theory are in the limit of
large 't Hooft coupling 
$\lambda=g_{YM}^2 \rN$ and large N dual to supergravity theories on an 
AdS$_5$-background.

In particular the Maldacena conjecture in the large N limit has been 
generalised to the non-supersymmetric case by placing D3-branes on 
non-supersymmetric singularities \refs{\rsilver,\rlnv}. 
Generically, without supersymmetry on the string theory side
one expects a dilaton potential to be generated by string-loop corrections,
so that only in the large N limit one can expect the corresponding
non-supersymmetric gauge theories to be conformal. For finite N one gets
1/N corrections and one still hopes that the exact $\beta$ functions 
allow at least one interacting fixed point. In a recent proposal \rframp\ such
non-supersymmetric conformal gauge theories were advertised as a
different scenario to solve the hierarchy problem of scales. Instead
of a supersymmetric gauge theory broken in the TeV range, one might
use a non-supersymmetric gauge theory which is conformal above some scale
close to the weak scale. In \rcsaki\ it was claimed that the 
one loop quadratic divergence of the Higgs mass does not cancel
in subleading orders in N.   
  
A different approach towards  non-supersymmetric gauge theories was 
pushed forward by Klebanov and Tseytlin
\refs{\rkletsya,\rkletsyb,\rkletsyc} following an idea by Polyakov \rpoly.
They studied the ten-dimensional tachyonic \typeo\ theory and found 
the remarkable result that non-trivial RR-flux can remove the
tachyonic instability. They studied both the case of parallel
electric D3-branes and the case of parallel self-dual D3-branes.
In the former case one gets a SU(N) gauge theory, with six adjoint scalars,
in which the one-loop renormalization group flow of the gauge coupling
is correctly captured by the non-constant dilaton in the
\typeo\ gravity solution. In the case of self-dual D3-branes,
in the limit of large N and $\lambda<100$ they found a stable AdS$_5\times
S^5$ solution, implying that in the large N limit the gauge theory is
conformal. This was further supported  by an explicit computation of the
two loop $\beta$-function for the gauge coupling. It vanished in the
large N limit but contained the expected non-zero 1/N corrections.
In \rnekr\ it was pointed out that this model can also 
be obtained as a $\ZZ_2$ orbifold of \typeb\ and thus is a member of 
the class of non-supersymmetric models studied in \refs{\rsilver,\rlnv}. 
Therefore, the general arguments presented
in \rbkv\ were applicable implying that the non-supersymmetric gauge
theory in question is conformal to all orders in the large N limit.  
Meanwhile, there have been discussions of several related issues of 
type 0 string theories such as the infrared behaviour \rminahan,
the physics of D-branes \refs{\rgarousi,\rzarembo,\rcosta,\rtsz}, 
branes on conifold singularities \roz, orientifolds
of \typeo\ \refs{\rbergab,\rangel,\rbfl}  and dualities \rbergman. 

In this paper we will systematically study non-supersymmetric gauge theories 
arising on self-dual D3-branes sitting on various singularities
in \typeo. In the case of supersymmetric singularities, the 
non-supersymmetric theories have some common features
with their corresponding supersymmetric \typeb\ cousins. Since the open
string annulus amplitude for self-dual D3-branes is up to factor of two the
same as the corresponding \typeb\ amplitude, all models have the same
number of bosons and fermions. In this sense they are special and
similar to the compact non-supersymmetric models discussed in \rkachru.
Using the arguments given in \rbkv, 
in the large N limit all correlation functions of the \typeo\ theories
reduce to the  correlation functions of the parent \cN=4 gauge theory. 
Thus, the \typeo\ orbifold theories are conformal in the large N  limit. 
Actually, it turns out that they have the same one-loop $\beta$-function as 
their  supersymmetric cousins and the first 1/N corrections appear 
at two loops. As we mentioned before, these \typeo\ gauge theories 
are a very special subset of large N non-supersymmetric conformal gauge 
theories and inherit some nice features from the corresponding supersymmetric
gauge theories. Due to this, we think they deserve particular interest.

As was pointed out in \refs{\rsagbi,\rsagn} and further elaborated in 
\refs{\rbergab,\rangel,\rbfl}
there exist different orientifold projections in \typeo. We will
also study  effective gauge theories on self-dual D3-branes on
orientifolds. In contrast to the pure orbifold case, generically there
remain some uncancelled twisted NSNS tadpoles, which are however suppressed
in the large N limit. Thus we can still construct large N conformal
gauge theories now including orthogonal and symplectic gauge groups and
also matter in the (anti-)symmetric representation of the unitary gauge
factors. The corresponding supersymmetric cases were discussed in
\refs{\rkakor,\rparur}. 

For the supersymmetric case, it is well established that D3-branes
on ALE type singularities are T-dual to cyclic Hanany-Witten 
models \refs{\rkarch,\rhanur,\randreas,\ruranga,\rdasmuk,\ragana}. 
Using the low energy effective action derived in \rkletsya, we will
argue that the same holds for self-dual D3-branes in \typeo. 
Essentially, since in  the effective theory for self-dual D3-branes 
the tachyon has the expectation value $\la T\ra=0$, as long as
the solution is stable the theory reduces
to the \typeb\ effective theory and everything carries over.  

Thus, using NS 5-branes and dyonic D-branes we  construct the type 0
analogue of Hanany-Witten models. This can be done for  non-conformal cases, 
as well.  We will discuss the ``\cN=2'' Hanany-Witten models in some
more detail, including the determination of the massless spectrum and
some comments about the moduli space.  
The ultimate goal would of course be, to gain some higher loop
or non-perturbative information using the recently proposed duality
of type 0A to M-theory on $S^1/(-1)^{F_s} S$ \rbergman.

\newsec{Branes on orbifold singularities in Type 0B}

In this section we study non-supersymmetric gauge theories obtained
by placing D3-branes on orbifold singularities \rdougmoor\ in Type 0B string
theory.
As was already observed in \rkletsyb, in order to obtain a large N conformal
theory one needs to have an equal  number of electric and magnetic 
D3-branes. If we did not take  the same number of electric and magnetic
D-branes the annulus amplitude would be non-vanishing implying a
non-trivial dilaton potential. In the field theory this would lead to 
a non-vanishing one loop $\beta$ function even in the large N limit.
To begin with we briefly review the definition of Type 0B and  the parent
gauge theory living on parallel D3-branes.

\subsec{Type 0B string theory}

There are two ways of constructing \typeo\ string theory. One is by
implementing in the superstring the projection  
$P={1\over 2}(1+(-1)^{F_L+F_R})$ \rseibwit. 
In contrast to the usual GSO projection
the tachyon survives and  all space-time fermionic modes are projected
out. The second way of constructing Type 0B is by an orbifold
of \typeb\ by the space-time fermion number $(-1)^{F_s}$. This removes
all fermions from the untwisted sector and leads to a tachyon and further
massless RR fields in the twisted sector. 
Computing the spectrum one finds that all states in the RR sector are 
doubled implying that all Dp-branes (p odd) are doubled, as well. 
In the case of D3-branes now there exist electric (D3) and magnetic (D$3'$)
branes.
Using the boundary state approach it was shown in \rbergab\ that indeed the 
boundary
state representing a Dp brane in \typeb\ splits into two boundary states
of \typeo. The explicit form of the boundary states was used to derive
rules for open strings stretched between the various types of Dp branes.
Open strings stretched between the same kind of Dp branes carry only
space-time bosonic modes, whereas open strings stretched between a D3 and
a D$3'$-brane carry only space-time fermionic modes. \pano
Now we consider \typeo\ as the orbifold of \typeb\ by $(-1)^{F_s}$ and
compute the annulus amplitude for 2N D3-branes in the loop channel
\eqn\cyla{ A= \int_0^\infty \, {{dt}\over t} \, 
                 {\rm Tr}_{open} \left[ \textstyle{{(1+(-1)^f)\over 2}\, 
                 {(1+(-1)^{F_s})\over 2}}\  e^{-2\pi t L_0} \right]. }
We allow for a non-trivial action of $(-1)^{F_s}$ on the Chan-Paton
factors parametrised by the $\gamma$ matrix\footnote{$^{1}$}{This 
is the string theoretic analogue of the construction made in \rnekr\
in the context of the gauge theories.}
\eqn\gammaa{ \gamma_{(-1)^{F_s}}=\left(\matrix{  1_{n,n}  &  0  \cr
                                        0 & -1_{m,m} \cr} \right)_{2N,2N} }
with $n+m=2N$. 
For general $n$ and $m$ one obtains
\eqn\cylb{\eqalign{ A={V_4\over (8\pi^2 \a')^2} \int_0^\infty \, 
      {{dt}\over t^3} \,
       &{1\over 2} ({\rm Tr}\,\gamma_1)^2 \left(
      {f_3^8(e^{-\pi t})-f_4^8(e^{-\pi t})-f_2^8(e^{-\pi t})
                      \over f_1^8(e^{-\pi t}) } \right) + \cr
       &{1\over 2} ({\rm Tr}\,\gamma_{(-1)^{F_s}})^2 \left(
      {f_3^8(e^{-\pi t})-f_4^8(e^{-\pi t})+f_2^8(e^{-\pi t})
                      \over f_1^8(e^{-\pi t}) } \right) .}}
We would like to discuss two choices of $m$ and $n$ in more detail.
\item{a.)} For $n=m=N$ \cylb\ reduces to
\eqn\cylc{ A={V_4\over (8\pi^2 \a')^2} \int_0^\infty \, {{dt}\over t^3} \,
       {2} N^2 \left(
      {f_3^8(e^{-\pi t})-f_4^8(e^{-\pi t})-f_2^8(e^{-\pi t})
                      \over f_1^8(e^{-\pi t}) } \right) }
which is exactly the amplitude for the same number
of electric and magnetic D3-branes in \typeo. Up to a factor of two this
is identical with the annulus amplitude for N D3-branes in \typeb. 
In particular, even without supersymmetry the amplitude vanishes and
one does not get any tachyonic tadpole.  \smno
\item{b.)} For $n=2N$ and $m=0$ \cylb\ reduces to
\eqn\cyld{ A={V_4\over (8\pi^2 \a')^2} \int_0^\infty \, {{dt}\over t^3} \,
        (2N)^2 \left(
      {f_3^8(e^{-\pi t})-f_4^8(e^{-\pi t}))
                      \over f_1^8(e^{-\pi t}) } \right), }
which is nothing else than the amplitude for 2N electric D3-branes. As we 
mentioned above in this case one only gets space-time bosonic modes
and a tachyonic tadpole in the tree channel.  \smno
From these two examples we learn that the numbers $m$ and $n$ in the
$\gamma_{(-1)^{F_s}}$ matrix are exactly the number of D3 and D$3'$-branes
in \typeo. The massless spectra living on the D3-branes in the two 
cases a.) and b.) are
\item{a.)} Gauge group $G={\rm SU(N)}\times {\rm SU(N)}$
           \footnote{$^2$}{As usual  the abelian U(1) subgroups 
             decouple in the infrared}, 
      three complex bosons in the
           adjoint and four Weyl-fermions in the $(\rN,\o{\rN})+(\o{\rN},\rN)$
           representation of $G$.\pano
\item{b.)} Gauge group $G={\rm SU(2N)}$ with three complex bosons in the
           adjoint representation of SU(2N). \pano
In case a.) the one-loop $\beta$-function vanishes and as was shown explicitly
in \rkletsyb\  the two-loop $\beta$-function vanishes only in the large 
N limit. 
As was already observed in \rnekr, since we have embedded the gauge theory
into \typeb\ string theory with a $\gamma$ matrix of vanishing trace, the
general arguments of \rbkv\ tell us that in the large N limit the 
non-supersymmetric SU(N)$\times$SU(N) gauge theory is conformal.
In the rest of this section we present the massless spectra obtained
by putting self-dual D3-branes on singularities preserving \cN=2, \cN=1 and
\cN=0 supersymmetry, respectively. 

\subsec{D3-branes on \cN=2, $\ZZ_K$ singularities}

We are considering the $\ZZ_K$ singularity described by the action
\eqn\singa{ \Theta=\cases{  z_1\to \theta z_1 \cr
                     z_2\to \theta^{-1}  z_2 & with 
                                      $\theta=e^{2\pi i /K}$ \cr
                     z_3\to z_3 \cr} }
on the three complex coordinates transversal to  $M$ D3-branes.
This case has also been discussed in \rbillo. Note, that since we are
considering non-compact models there is no need to restrict
to $\ZZ_K$ actions having a crystallographic action on the $T^4$ torus and
therefore every non-negative integer $K$ is allowed.  
As we have discussed  in the last subsection, the annulus amplitudes for 
self-dual D3-branes in \typeo\ are up to factor of two identical 
to the annulus amplitudes for D3-branes in \typeb. This means that
the twisted tadpole cancellation conditions are the same, as well. 
Moreover, it means that the \typeo\ spectrum is bose-fermi degenerated. 
As was shown in \rleigh, since $\Theta$ leaves one coordinate invariant,
all twisted  tadpoles are of logarithmic type
\eqn\singb{   \sum_{k=1}^{K-1}\, \int_0^\infty {dl\over l} 
             {\rm Tr}(\gamma_{\Theta^k}){\rm Tr}(\gamma^{-1}_{\Theta^k}) .}
These tadpoles are  in one to one correspondence with the vanishing of the
one-loop $\beta$ function of the gauge couplings.  
Even without cancelling them, one gets an anomaly free gauge theory. 
We will see that the same holds for the \typeo\ case. \pano
By choosing
\eqn\singc{ \gamma_\Theta=
   {\rm diag}[1_{n_1}, \theta_{n_2},\ldots,\theta^{K-1}_{n_K}] }
with $\sum n_j=M=NK$, in the \cN=2 supersymmetric case the massless 
spectrum on the D3-branes is
\eqn\singd{\eqalign{ &{\rm vectormultiplets:}\quad G=\bigotimes_{j=1}^K 
                                             {\rm SU}(n_j) \cr
                     &{\rm hypermultiplets:}\quad \bigoplus_{j=1}^K\, 
                 (\funda_j,   \bfunda_{j+1} ) }}
with cyclic identification. 
Making the choice 
\eqn\singe{  \Gamma_\Theta=\left(\matrix{  \gamma_\Theta  &  0  \cr
                                   0  & \gamma_\Theta \cr} \right)_{2M,2M} }
for the action of $\ZZ_K$ on the Chan-Paton factors of the D3 and D$3'$-branes
one derives the following gauge group for the \typeo\ model
\eqn\singf{ G=\bigotimes_{j=1}^K {\rm SU}(n_j) \times 
                       \bigotimes_{j=1}^K {\rm SU}(n_j) }
where the first K factors arise on the D3-branes and the second K factors
on the D$3'$-branes. There are additional  complex bosons and Weyl-fermions 
in the following representations
\eqn\singg{\eqalign{ 
         \bigoplus_{j=1}^K  
        &\left\{  \zwei{{\bf\rm Adj}_j}{1} + \zwei{1}{{\bf\rm Adj}_j}+
                 2\vier{\funda_j}{\bfunda_{j+1}}{1}{1}+
                 2\vier{1}{1}{\funda_j}{\bfunda_{j+1}}
                 \right\}_B + \cr
        \bigoplus_{j=1}^K 
        &\biggl\{ 2\zwei{\funda_j}{\bfunda_j} + 
                2\zwei{\bfunda_j}{\funda_j} +
                \vier{\funda_j}{1}{1}{\bfunda_{j+1}}+
                \vier{\bfunda_j}{1}{1}{\funda_{j+1}}+ \cr
                &\vier{1}{\funda_{j+1}}{\bfunda_{j}}{1}+
                \vier{1}{\bfunda_{j+1}}{\funda_{j}}{1}
                 \biggr\}_F \cr }}
where the upper row in the matrix notation refers to the gauge groups
on the D3-branes and the lower row refers to the gauge groups 
on the D$3'$-branes. From \singg\ it is evident that even without 
cancelling the tadpoles
the non-chiral  spectrum is free of non-abelian gauge anomalies. 
Requiring cancellation of the logarithmic tadpoles leads to $n_j=M/K$ 
for all $j\in\{1,\ldots,K\}$ implying indeed that the one-loop 
$\beta$-function vanishes. 

\subsec{D3-branes on \cN=1, $\ZZ_K$ singularities}

We are considering the $\ZZ_K$ singularity described by the action
\eqn\singh{ \Theta=\cases{  z_1\to \theta^{l_1} z_1 \cr
                     z_2\to \theta^{l_2}  z_2 \cr
                     z_3\to \theta^{l_3} z_3 \cr} }
with $l_1+l_2+l_3=0$.
The \cN=1 supersymmetric spectrum of these models is
\eqn\singi{\eqalign{ &{\rm vector:}\quad G=\bigotimes_{j=1}^K 
                                             {\rm SU}(n_j) \cr
                     &{\rm chiral:}\quad  \bigoplus_{a=1}^3\,
                    \bigoplus_{j=1}^K\, (\funda_j,\bfunda_{j+l_a}) ,}}
which in general is chiral. Choosing the $\gamma$ matrices as in \singe\
leads to the following non-supersymmetric spectrum for the \typeo\ case.
\eqn\singj{\eqalign{ &G=\bigotimes_{j=1}^K {\rm SU}(n_j) \times 
                       \bigotimes_{j=1}^K {\rm SU}(n_j) \cr
         &\bigoplus_{j=1}^K \bigoplus_{a=1}^3   
        \left\{  \vier{\funda_j}{\bfunda_{j+l_a}}{1}{1}+
                 \vier{1}{1}{\funda_j}{\bfunda_{j+l_a}}
                 \right\}_B + \cr
        &\bigoplus_{j=1}^K  
        \left\{ \zwei{\funda_j}{\bfunda_j} + 
                \zwei{\bfunda_j}{\funda_j} + \bigoplus_{a=1}^3 \left[
                \vier{\funda_j}{1}{1}{\bfunda_{j+l_a}}+
                \vier{1}{\bfunda_{j+l_a}}{\funda_{j}}{1} \right]
                 \right\}_F .\cr }}
The non-abelian gauge anomaly of SU($n_j$) is proportional to
\eqn\singk{ \sum_{a=1}^3 (n_{j+l_a}-n_{j-l_a})}
and the one-loop $\beta$ function of the corresponding gauge coupling is  
\eqn\singk{ \beta_1=3 n_j -{1\over 2} \sum_{a=1}^3 (n_{j+l_a}+n_{j-l_a}),}
which agrees with the data for the \cN=1 supersymmetric model in \singi.
Both the anomalies and the  one-loop $\beta$ functions vanish if all tadpoles
are cancelled for $n_j=N$.      

\subsec{D3-branes on \cN=0, $\ZZ_K$ singularities}
                       
Since we are interested in models without supersymmetry there is no need 
to consider supersymmetric singularities only.  However, to have a well
defined string description one has to guarantee that the model satisfies
level matching. Let us examine one specific example, where we first
consider the \typeb\ orbifold case. 
We start with the $\ZZ_2$ action $\Theta:z_i\to -z_i$ 
for all $i\in\{1,2,3\}$.
Checking for level matching one realizes that one gets $\Delta E={1\over 4}$
which cannot be compensated in  a $\ZZ_2$ orbifold.
Indeed, since $K=2$ and $l_i=1/2$, $\Theta$ does not satisfy the
consistency condition $K(l_1+ l_2 + l_3) ={\rm even}$.
If one would naively continue and compute the  open string sector
one would indeed find that in the  Ramond sector $\Theta$ acts like
a $\ZZ_4$, as well. However, it
is possible to consider the orbifold action above as a $\ZZ_4$
orbifold, $\vec l={1\over 4}(2,2,2)$, with partition function
\eqn\singk{ Z={1\over 2}\left(  {\scriptstyle{1}}\square\limits_1+
                   {\scriptstyle{\Theta}}\square\limits_1 \right) +
              {1\over 4} \left( {\scriptstyle{1}}\square\limits_{\Theta}+
              {\scriptstyle{\Theta}}\square\limits_{\Theta}+
              {\scriptstyle{\Theta^2}}\square\limits_{\Theta}+
              {\scriptstyle{\Theta^3}}\square\limits_{\Theta} \right) .}
Requiring tadpole cancellation leads to the $\gamma$ matrix
\eqn\singl{ \gamma_{\Theta}={\rm diag}[1_N,i\, 1_N,-1_N,-i\, 1_N] }
satisfying $\gamma_{\Theta}^4=1$. For the D3-brane spectrum we
obtain
\eqn\singi{\eqalign{ {\rm vector:}\quad &G=\bigotimes_{j=1}^4 
                                             {\rm SU(N)} \cr
                     {\rm matter:}\quad  &\bigoplus_{j=1}^2\,
                                3 (\funda_j,\bfunda_{j+2})_B +
                     \bigoplus_{j=1}^4\,
                                4 (\funda_j,\bfunda_{j+1})_F   .}}
The spectrum is free of non-abelian gauge anomalies and has vanishing
one-loop $\beta$ function. This chiral model was also derived in 
\rframp\  using the approach of \rlnv. 
The corresponding \typeo\ model is
\eqn\singj{\eqalign{ &G=\bigotimes_{j=1}^4 {\rm SU(N)} \times 
                       \bigotimes_{j=1}^4 {\rm SU(N)} \cr
         &\bigoplus_{j=1}^2    
        \left\{  3\vier{\funda_j}{\bfunda_{j+2}}{1}{1}+
                 3\vier{1}{1}{\funda_j}{\bfunda_{j+2}}
                 \right\}_B +\cr
        &\bigoplus_{j=1}^4  
        \left\{ 4\vier{\funda_j}{1}{1}{\bfunda_{j+1}}+
                4\vier{1}{\bfunda_{j+1}}{\funda_{j}}{1} 
                 \right\}_F ,\cr }}
which is free of non-abelian gauge anomalies and has vanishing
one-loop $\beta$ function, as well. 
So far all the \typeo\ spectra we derived had only an even number of
unitary gauge groups and
bifundamental matter. The generalisation to orthogonal and symplectic
gauge  groups and an odd number of unitary gauge factors 
is achieved by using \typeo\ orientifolds.

\newsec{Non-susy gauge theories from orientifolds of Type 0B}
\noindent
As was first pointed out in \rsagbi\ and elaborated further in 
\refs{\rsagn,\rbergab,\rangel,\rbfl,\rbergman}, in the
ten-dimensional \typeo\ string theory there exist three different
orientifold projections. 

\item{i.)} First, one has the usual world-sheet
parity transformation $\Omega$, which in ten-dimensions leads to a
model containing tachyons both in the closed 
and in the open string sector. The tachyon in the open string 
sector is due to the untwisted tadpole cancellation
condition requiring the introduction of D9 and anti D9-branes.
We will see that due to the weaker tadpole cancellation conditions
in the non-compact case, one can get rid of the open string tachyons
leading to sensible gauge theories on the D3-branes including orthogonal
and symplectic gauge groups. In order to avoid tachyonic tadpoles
in the annulus amplitude, we will always work with self-dual D3-branes 
in the following. \smno

\item{ii.)} As was shown in \rbfl, the dressed 
parity transformation $\Omega'=\Omega\, (-1)^{f_R}$ projects out
the ten-dimensional closed string tachyon and does lead to a model
necessarily containing the same number of D9 and D$9'$-branes.  
The latter point is
simply an implication of the fact that the world-sheet fermion number
operator $(-1)^{f_R}$ exchanges Dp and Dp$'$-branes. In the non-compact case
such models also contain only self-dual D3-branes and 
generically lead to an  odd number of unitary gauge factors 
and  some fermions in the antisymmetric representation.  \smno

\item{iii.)} The third kind of ten-dimensional 
orientifolds of \typeo\ uses the 
combination $\Omega''=\Omega\, (-1)^{F_L}$ and was discussed in \rbergman. 
Since $(-1)^{F_L}$ exchanges D-branes with the corresponding anti D-brane, 
these models always contain open string tachyonic modes and 
thus they do not seem to be suitable for constructing large N 
conformal gauge theories. \smno
More specifically, we are interested in orientifolds by 
$\omega=\Omega (-1)^{F_L} J$ and 
$\omega'=\Omega' (-1)^{F_L} J$, respectively,  
where $J$ denotes  the $\ZZ_2$ transformation $z_i\to -z_i$ for all
$i\in\{1,2,3\}$ and $F_L$ is the left-moving space time fermion number 
operator. Note, that $\omega$ and $\omega'$ are T-dual to the ten-dimensional
$\Omega$ and $\Omega'$ discussed in i.) and ii.) above.
Besides the world sheet  parity transformation we are also gauging 
a further discrete space time symmetry $\ZZ_K$  acting as
$\Theta={1\over K}(v_1,v_2,v_3)$ on the three transversal coordinates.
The Klein bottle amplitude 
\eqn\oria{ K_{\omega'/\omega}=\int_0^\infty {dt\over t} {1\over K}
   \sum_{k=0}^{K-1}{\rm Tr}
  \left( \omega'/\omega \ \Theta^k \, {1+(-1)^{f_L+f_R} \over 2}
    e^{-2\pi t(L_0+\o L_0)} \right) }  
receives only contributions from the untwisted closed string
sector and for even $K$ also from the $\Theta^{K/2}$ twisted sector. 
\bigno
{\it The $\omega'$ orientifold}
\meno
Since $\omega'$ contains $(-1)^{f_R}$
the Klein bottle amplitude for \typeo\ string theory is up to factor
of two the truncation of the \typeb\ Klein bottle amplitude to terms
containing the insertion $(-1)^{f_R}$ in the trace. However, these are
exactly the terms leading to RR exchange in the tree channel interpretation
of the amplitude. 
The actual computation of \oria\ is identical to 
the \typeb\ computations presented in \refs{\riban,\rparur}.\pano  
In order to cancel the RR tadpoles in the $\omega'$ orientifold
one introduces D3/D$3'$-branes and
in case of even $K$ also D7/D$7'$-branes in the background. In the following
we will restrict ourselves to the $K$ odd case, the generalisation to
cases including D7/D$7'$-branes is straightforward. 
As we have reviewed in the section 2, the cylinder amplitude for
the same number of D3 and D$3'$-branes is exactly twice the amplitude
for D3-branes in \typeb\ and in particular it vanishes. 
Moreover, since $\omega'$ exchanges D3 with D$3'$-branes, the only non-vanishing
contribution to the M\"obius strip amplitude arises from open strings
stretched between D3 and D$3'$-branes. This again is exactly twice the
Ramond  part of the M\"obius amplitude for D3-branes in \typeb. Note,
that unlike the annulus amplitude the M\"obius amplitude is non-zero
and thus spoils the Bose-Fermi degeneracy.  
Thus, the tadpole cancellation conditions for RR exchange are identical
to the corresponding conditions in the analogous \typeb\ orientifold.
However there always remains an uncancelled twisted NSNS tadpole, which in the
string context can be cancelled by the Fischler/Susskind mechanism \rfisus.

The twisted RR tadpole cancellation conditions for the corresponding \typeb\ 
orientifolds have been derived in \refs{\riban, \rparur}
\eqn\orib{ {\rm Tr}(\gamma_{\Theta^{2k}})=\pm {1\over \prod_{i=1}^3\,
           {\cos}\left( {\pi kv_i\over K}\right) } }
where the two different signs refer to the two possible actions of
$\omega'$ on the Chan-Paton factors, namely whether  
$\gamma_{\omega'}$ is
antisymmetric or symmetric. Remember, that in the compact case this 
ambiguity is fixed by the untwisted tadpole cancellation condition. 
As shown in \riban, condition \orib\ always has a solution.
Taking into account that $\Omega'$ exchanges D3 and D$3'$-branes a consistent
choice of $\Gamma$ matrices in the \typeo\ orientifold  in terms
of the corresponding $\gamma$ matrices in the \typeb\ orientifold is
\eqn\oric{ \Gamma_\Theta=\left(\matrix{  \gamma_\Theta  &  0  \cr
                                   0  & \gamma_\Theta \cr} \right)_{2M,2M},
         \quad\quad
         \Gamma_{\omega'}=\left(\matrix{  
               0 & \gamma_{\omega}   \cr
              \gamma_{\omega} & 0 \cr} \right)_{2M,2M}, }
where $M$ denotes the number of self-dual D3-branes.
Note, that due to the tadpole cancellation condition \orib\ 
for $k\ge 1$ ${\rm Tr}(\gamma_{\Theta^{2k}})$ is of order O(N$^0$), whereas  
${\rm Tr}(\gamma_1)$ is of order O(N). Thus, one expects the effect of
the remaining
twisted NSNS tadpoles to be suppressed in the large N limit. 
We will indeed find, that opposed to the models in section 2 
the one-loop $\beta$ function of the gauge coupling receives 1/N corrections. 
\bigno
{\it The $\omega$ orientifold}
\meno
The story for the $\omega$ orientifold is a bit different.
The Klein bottle amplitude with
the $\omega$ insertion in \oria\ only contains terms leading to NSNS exchange
in tree channel. 
Here, we also
introduce the same number of D3 and D$3'$-branes, but in contrast to the
compact case there is no untwisted RR tadpole and therefore no need
to introduce anti-branes, as well. Thus, we do not have to worry about
extra   tachyons. The M\"obius amplitude contains only
traces over the NS sector leading to terms in the NSNS exchange channel. 
Summarising, only the annulus amplitude contains twisted RR tadpoles, whereas
all three amplitudes contribute to twisted NSNS tadpoles. Thus, in
order to cancel the dangerous RR tadpoles one has to require
\eqn\oricc{ {\rm Tr}(\gamma_{\Theta^{k}})=0 }
and like the $\omega'$ case one is left with an uncancelled 
massless twisted NSNS tadpole. 
A consistent choice of $\gamma$ matrices in the $\omega$  orientifold is
\eqn\oriop{ \Gamma_\Theta=\left(\matrix{  \gamma_\Theta  &  0  \cr
                                   0  & \gamma_\Theta \cr} \right)_{2M,2M},
         \quad\quad
         \Gamma_{\omega}=\left(\matrix{  
               \gamma_{\omega} & 0    \cr
               0 &  \gamma_{\omega} \cr} \right)_{2M,2M}. }
In the following we discuss a couple of examples of both $\omega$ and
$\omega'$ orientifolds. 

\subsec{Orientifolds on \cN=4 singularities}

The simplest model is taking the brane configuration from section 2.1
and gauge the world-sheet parity transformation. 
The corresponding \typeb\ model is \cN=4 susy SO(M) or SP(2M) gauge 
theory.  However, due to the different action of $\omega'$ on the
Chan-Paton factors in the \typeo\ model one obtains the following
gauge group
\eqn\orid{ G={\rm SU(N)}}
equipped with the bosonic and fermionic matter
\eqn\orida{3({\rm Adj})_B + 4\left( \asymm + \basymm \right)_F .}
For symmetric $\gamma_{\Omega'}$ one gets the symmetric
representation in \orida\ instead of the antisymmetric. 
Computing the one-loop $\beta$ function coefficient one realizes that it 
vanishes only in the large N limit
\eqn\orie{ b_1=0\, \rN \pm {16\over 3}. } 
Note, that for the antisymmetric representation the model is 
asymptotically free. 
The general form of the two-loop gauge coupling  coefficient 
$\beta$-function has been derived in \rmv\
\eqn\betaa{ b_2={34\over 3} C_2(G)^2 -\sum_S \left[ 2 C_2(S)+{1\over 3}
C_2(G)\right] T_2(S) -2 \sum_F \left[ C_2(F)+{5\over 3}
C_2(G)\right] T_2(F) + Y_4 ,}
where $C_2(R)$ denotes the eigenvalue of the second order Casimir in the
representation $R$ and $T_2(R)$ denotes the Dynkin-Index of $R$. 
The contribution of the Yukawa couplings is given by
\eqn\betab{ Y_4={1\over {\rm dim}(G)} {\rm Tr}\left[ \hat C_2(F) Y_I Y_I^*
\right] }
where $Y_I$ denotes the matrix of the Yukawa couplings.
Taking into account the global SU(4) symmetry and the gauge 
structure of $Y_I$, as in
\rkletsyb, leads to $Y_4= 24(\rN^2-3\rN +4/\rN)$. 
Hence, the $\rN^2$ term vanishes. 
Up to order $g^7$ the $\beta$ function is given by
\eqn\betac{\eqalign{ \beta&=-{g^3\over (4\pi)^2} b_1 -
              {g^5\over (4\pi)^4} b_2 \cr
               &=\mp {g^3\over (4\pi)^2}{16\over 3} \pm {g^5\over (4\pi)^4}
              {64\over 3} \left( \rN-{3\over \rN} \right) + O(g^7) .}}
For $\rN\ge 2$ the coefficients $b_2$ and $b_1$ have  opposite signs, 
which at least
supports the hope that there exists an interacting fixed-point for some finite
value of the coupling constant $g$. Unfortunately, \betac\ implies 
$g_*^2=O(1/\rN)$
telling us that for large but finite N higher loop contributions to the
$\beta$ function can not be neglected.  
\pano
Taking instead the  $\omega$ projection one obtains the gauge group
\eqn\soa{ G={\rm SO(N)}\times {\rm SO(N)}}
with matter
\eqn\sob{3\left\{({\rm Adj},1)+(1,{\rm Adj})\right\}_B + 
         4 ( \funda,\bfunda )_F .}
Choosing  $\gamma_{\Omega}$ symmetric leads to  SP(2N) gauge groups.
For the  $\beta$ function up to two loops one obtains
\eqn\socb{\eqalign{ \beta&=-{g^3\over (4\pi)^2} b_1 -
              {g^5\over (4\pi)^4} b_2 \cr
               &=\pm {g^3\over (4\pi)^2}{16\over 3} \mp {g^5\over (4\pi)^4}
              {64\over 3} \left( \rN\mp {1\over 2} \right) + O(g^7) ,}}
where again the highest order terms vanish and $b_2$ and $b_1$ have opposite
sign.\pano
In contrast to section 2, in all \typeo\ orientifold models we studied 
the one-loop $\beta$ function 
vanished only in the large N limit. 
The string theoretic reason for the running of the gauge coupling is, that
in contrast to the annulus amplitude the M\"obius amplitude does
not vanish. This one-loop cosmological constant induces a dilaton potential
which prevents the dilaton from being a free parameter. Since, the M\"obius
amplitude is 1/N suppressed against the annulus amplitude, in the field
theory this scenario induces a 1/N correction to
the one-loop $\beta$ function.

\subsec{Orientifolds on \cN=2 singularities}

The spectrum for \typeb\ orientifold of the models presented in section
2.2 for odd $K=2P+1$  was derived in \rparur
\eqn\orif{\eqalign{ &{\rm vectormultiplets:}\ \ G={\rm SO(N)}\times 
                           {\rm SU(N-2)}\times
                                   \dots \times {\rm SU(N-2P)}  \cr
                     &{\rm hypermultiplets:}\ \  \bigoplus_{j=1}^{P-1}\, 
                 (\funda_j,   \bfunda_{j+1} ) +\asymm_P  .\cr }} 
In the corresponding $\omega'$  orientifold the SO(N) group becomes
an SU(N) gauge group and all the remaining SU$(n_j)$ groups with
$n_j=\rN-2j$ are doubled. 
The complete gauge group is
\eqn\orig{ G={\rm SU(N)}\times \bigotimes_{j=1}^P {\rm SU}(n_j) \times 
                       \bigotimes_{j=1}^P {\rm SU}(n_j) }
where the first $P$ factors arise on the D3-branes and the second $P$ factors
on the D$3'$-branes. The bosonic and fermionic matter is given by
\eqn\orih{\eqalign{ 
      &\left\{ ({\bf\rm Adj}_0) + \zwei{{\bf\rm Adj}_P}{1} + 
                \zwei{1}{{\bf\rm Adj}_P}+
                   2\drei{\funda_0}{\bfunda_{1}}{1}+
                 2\drei{\funda_0}{1}{\bfunda_{1}}+
                 2\zwei{\funda_P}{\bfunda_P} \right\}_B + \cr
       &    \bigoplus_{j=1}^{P-1} \left\{
                 \zwei{{\bf\rm Adj}_j}{1} + \zwei{1}{{\bf\rm Adj}_j}+
                 2\vier{\funda_j}{\bfunda_{j+1}}{1}{1}+
                 2\vier{1}{1}{\funda_j}{\bfunda_{j+1}} 
                 \right\}_B +\cr
        &\biggl\{2\,\asymm_0+2\,\basymm_0+2\zwei{\funda_P}{\funda_P}+
               2\zwei{\bfunda_P}{\bfunda_P}+
               \drei{\funda_0}{\funda_{1}}{1}+
               \drei{\bfunda_0}{\bfunda_{1}}{1}+ 
               \drei{\funda_0}{1}{\funda_{1}}+ \cr
             &\ \   \drei{\bfunda_0}{1}{\bfunda_{1}}+
               \zwei{\asymm_P}{1}+\zwei{\basymm_P}{1}+\zwei{1}{\asymm_P}+
               \zwei{1}{\basymm_P}
                 \biggr\}_F +\cr
           & \bigoplus_{j=1}^{P-1} \biggl\{
              2\zwei{\funda_j}{\funda_j}+
               2\zwei{\bfunda_j}{\bfunda_j}+
                \vier{\funda_j}{1}{1}{\funda_{j+1}}+
                \vier{\bfunda_j}{1}{1}{\bfunda_{j+1}}+ \cr
              &  \vier{1}{\funda_{j+1}}{\funda_{j}}{1}+
                \vier{1}{\bfunda_{j+1}}{\bfunda_{j}}{1}
                 \biggr\}_F. \cr }}
The one loop $\beta$ function coefficients are as follows 
\eqn\orii{ b_0={20\over 3},\ b_1=\ldots=b_{P-1}=0,\   b_P=-{2\over 3}. }
The spectrum for the other sign in \orib\ is
$G={\rm Sp(2N)}\times {\rm SU(2N+2)}\times \dots \times {\rm SU(2N+2P)}$ 
with the antisymmetric representations exchanged by the symmetric ones.  
We skip the representation of the corresponding $\omega$ orientifold.

\subsec{Orientifolds on \cN=1 singularities}

In this case one can not give such a closed form for  the spectra 
depending on the vector $\Theta={1\over K}(v_1,v_2,v_3)$ with 
$v_1+v_2+v_3=0$. 
However, it is a straightforward exercise to find the solution
of \orib\ and then use the explicit form of $\gamma_\Theta$
to determine the spectrum of the desired model. 
Here we would only like to discuss
the simplest model, which is the $\ZZ_3$ orientifold. 
The gauge group for the $\omega'$ orientifold is
\eqn\orij{ G={\rm SU(N-4)}\times {\rm SU(N)} \times  {\rm SU(N)} }
The bosonic and fermionic matter content is given by
\eqn\orik{\eqalign{ 
      &3\left\{ \drei{\funda}{\bfunda_{1}}{1} +
               \drei{\bfunda}{1}{\funda} + \drei{1}{\funda}{\bfunda}
             \right\}_B +\cr
        &3\left\{ \drei{1}{\asymm}{1} + \drei{1}{1}{\basymm}+
              \drei{\bfunda_0}{\bfunda}{1}+
               \drei{\funda_0}{1}{\funda_{1}} \right\}_F +  \cr
       &\left\{ \drei{\asymm}{1}{1} +\drei{\basymm}{1}{1} +
                \drei{1}{\funda}{\funda}+
                \drei{1}{\bfunda}{\bfunda} \right\}_F. }}
The model is chiral but free of non-abelian gauge anomalies. 
The one-loop $\beta$ function coefficients are $b_0=-32/3$ and $b_1=8$.\pano
Computing the gauge group for the $\omega$ orientifold 
\eqn\orijj{ G={\rm SO(N)}\times {\rm SU(N)} \times {\rm SO(N)}
               \times{\rm SU(N)} ,}
we would like to emphasise that due to string consistency namely RR tadpole
cancellation we get SO(N) gauge factors instead of the \typeb\ result
SO(N-4). The matter spectrum is
\eqn\orikk{\eqalign{ 
      3&\left\{ \vier{1}{\asymm}{1}{1}+\vier{1}{1}{1}{\asymm}+
               \vier{\funda}{\bfunda}{1}{1}+ 
                \vier{1}{1}{\funda}{\bfunda}\right\}_B +\cr
      &\left\{ \vier{\funda}{1}{\funda}{1}+\vier{1}{\funda}{1}{\bfunda}+
              \vier{1}{\bfunda}{1}{\funda} \right\}_F +\cr
      3&\left\{ \vier{\funda}{1}{1}{\bfunda}+\vier{1}{\bfunda}{1}{\funda}+
              \vier{1}{\funda}{\funda}{1} \right\}_F \cr }}
Note, the chiral matter contents in \orikk\ is free of non-abelian
gauge anomalies and  
the one-loop $\beta$ function coefficients
 are $b_0=b_2=\pm 22/3$ and $b_1=b_3=\pm 1$.

\subsec{Orientifolds on \cN=0 singularities}

In this section we will discuss orientifolds of both \typeb\ and 
\typeo\ on a non-supersymmetric $\ZZ_5$ singularity. The $\ZZ_5$ action
is defined as $\Theta={1\over 5}(0,1,3)$ and satisfies level matching. 
Solving for the twisted RR tadpole conditions yields
\eqn\oril{ \gamma_{\Theta}={\rm diag}[1_{N+2},\theta_N,\theta^2_N,\theta^3_N,
 \theta^4_N] .}
There are also tachyon and dilaton tadpoles in the NSNS exchange channel, 
however they are cancelled automatically since they always appear in
the same combination with the RR channel tadpoles.
One obtains the following gauge group  of the \typeb\ orientifold
\eqn\orim{ G={\rm SO(N+2)}\times {\rm SU(N)} \times  {\rm SU(N)} }
equipped with some bosonic and fermionic matter given by
\eqn\orin{\eqalign{ 
      &\left\{ \bigoplus_{j=1}^3 2({\bf\rm Adj}_j) +
               (\funda,\bfunda,1) + (\funda,1,\bfunda) + 2 (1,\funda,\bfunda)
               +(1,\asymm,1)+ (1,1,\asymm\,) \right\}_B +\cr
      &\left\{ (\funda,\bfunda,1) + (\funda,1,\bfunda) + 2 (1,\funda,\bfunda)
               +(1,\asymm,1)+ (1,1,\asymm\,) + c.c. \right\}_F .\cr }}
Since the $\ZZ_5$ leaves one of the coordinates invariant, the spectrum
is non-chiral and has vanishing one-loop $\beta$ function.
The derivation of the \typeo\ $\omega'$ orientifold is straightforward
and gives the the gauge group
\eqn\orio{ G={\rm SU(N+2)}\times {\rm SU(N)} \times  {\rm SU(N)} 
                          \times {\rm SU(N)} \times  {\rm SU(N)} }
with matter
\eqn\orip{\eqalign{
&\left\{ \funf{{\bf\rm Adj}}{1}{1}{1}{1} + {\rm cycl.} \right\}_B +\cr
&\left\{ \funf{\funda}{\bfunda}{1}{1}{1}+\funf{1}{\funda}{\bfunda}{1}{1}+
         \funf{1}{1}{\funda}{\bfunda}{1}+\funf{1}{1}{1}{\funda}{\bfunda}+
         \funf{\bfunda}{1}{1}{1}{\funda}  \right\}_B +\cr
&\left\{ \funf{\funda}{1}{\bfunda}{1}{1}+\funf{1}{\funda}{1}{\bfunda}{1}+
         \funf{1}{1}{\funda}{1}{\bfunda}+\funf{\bfunda}{1}{1}{\funda}{1}+
         \funf{1}{\bfunda}{1}{1}{\funda}  \right\}_B  +\cr
&\left\{ \funf{\funda}{\funda}{1}{1}{1}+\funf{1}{\funda}{1}{\funda}{1}{1}+
         \funf{1}{1}{\asymm}{1}{1}+\funf{1}{1}{1}{\asymm}{1}+
         \funf{\funda}{1}{1}{1}{\funda}+
         \funf{1}{1}{\funda}{1}{\funda}+ c.c.  \right\}_F +\cr
&\left\{ \funf{\funda}{1}{\funda}{1}{1}+\funf{1}{\funda}{\funda}{1}{1}{1}+
         \funf{1}{\asymm}{1}{1}{1}+\funf{1}{1}{1}{1}{\asymm}+
         \funf{\funda}{1}{1}{\funda}{1}+
         \funf{1}{1}{1}{\funda}{\funda}+ c.c.  \right\}_F. \cr }}
With this non-supersymmetric example we finish our discussion of \typeo\
orientifold models.

\newsec{D3-branes on \cN=1 conifold singularities in type 0B}

While for branes at orbifold singularities the spectrum can be
derived from perturbative string calculations, this is not anymore the case
placing  D3-branes at transversal conifold singularities. 
Here the spectrum has
to be determined by an educated `guess' work, which of course has to pass
several subsequent consistency requirements by considering various known
limits. The supersymmetric type IIB D3-branes at conifold singularities,
leading to  \cN=1 supersymmetric field theories, were first
discussed in \refs{\rkeha,\rklewit}. This discussion was then extended in 
\refs{\ruranga,\rdasmuk,\ragana}.
 We start with the 6-dimensional conifold singularity
${\cal C}$, which can be described by the following hypersurface in 
$\IC^4$:
\eqn\conifold{
{\cal C}:\quad xy=uv.
}
This case was recently studied in \roz.
As a first generalization  we consider 
orbifolds of the conifold singularity, namely the quotient ${\cal C}/\Gamma$, 
$\Gamma=\ZZ_K\times \ZZ_L$, i.e.
$x\rightarrow e^{2\pi i/K}x$, $y\rightarrow e^{-2\pi i/K}y$,
$u\rightarrow e^{2\pi i/L}u$, $v\rightarrow e^{-2\pi i/L}v$.
The corresponding hypersurface in $\IC^4$ is determined by the equation
\eqn\conifoldc{
{\cal C}_{KL}:\quad (xy)^L=(uv)^K.
}
This singularity is probed by $N$ `self-dual' D3-branes ($m=n=N$).
The spectrum can be obtained from 
the related \cN=1 supersymmetric models of D3-branes on these spaces 
in  type IIB 
superstrings \refs{\ruranga,\ragana} via the $(-1)^{F_s}$ projection.
The gauge group has
the following form
\eqn\conigauge{
G=\bigotimes_{j=1}^{KL}
\left({\rm SU(N)}_e\times {\rm SU(N)}_e\times {\rm SU(N)}_m
   \times {\rm SU(N)}_m\right).}
The subscript `$e$' (`$m$') stands here for electric (magnetic) gauge group
due to  the $N$ D3 (D$3'$) branes.   
We label each pair of electric plus magnetic group factor ${\rm SU(N)}_e\times
{\rm SU(N)}_m$ by indices $i,I=1\dots K$ and $j,J=1\dots L$.
Then we obtain four types of massless scalar and fermion matter fields
which are bifundamental under the gauge groups indicated by the indices:
\eqn\abfields{\eqalign{
\phi,\psi:\quad &(A_1)_{i+1,j+1;I,J}=(\funda_{i+1,j+1},\bfunda_{I,J}),\cr
&(A_2)_{i,j;I,J}=(\funda_{i,j},\bfunda_{I,J}),\cr
&(B_1)_{i,j;I,J+1}=(\funda_{i,j},\bfunda_{I,J+1}),\cr
&(B_2)_{i,j;I+1,J+1}=(\funda_{i,j},\bfunda_{I+1,J+1})
.\cr}}
The scalars are bi-fundamental under two electric or two magnetic gauge group
factors. On the other hand, 
with respect to the indices $e$ and $m$, the fermions are either in the
representations $(\funda_e,\bfunda_m)$ or $(\funda_m,\bfunda_e)$.
In addition there are also massless fermions which are bifundamental
under electric/magnetic gauge groups with identical set of indices:
\eqn\fermionsconi{
\psi:\quad (\funda_{i,j}^e,\bfunda_{i,j}^m)+(\funda_{i,j}^m,\bfunda_{i,j}^e)+
(\funda_{I,J}^e,\bfunda_{I,J}^m)+(\funda_{I,J}^m,\bfunda_{I,J}^e).}
The corresponding 1-loop $\beta$-function takes again the same value as in the
corresponding type IIB parent model.

Just as in the corresponding \cN=1 models, there exist a special Higgs
branch in this class of models
\refs{\ragana,\rtatar}. Specifically, giving a vev to all $A_2$
type scalar fields corresponds to resolving the original conifold singularity
to the well known orbifold singularity $\IC^3/(\ZZ_K\times \ZZ_L)$.
In this way all  the $A_2$ scalar fields will break each
${\rm SU(N)}_{ij}^{e,m}\times {\rm SU(N)}_{IJ}^{e,m}$ pair down to 
its diagonal subgroup. 
The remaining three types of massless fields after the Higgs mechanism
are precisely the horizontal, vertical and diagonal representations
which one gets on the $\ZZ_k\times\ZZ_L$ orbifold models.

Let us briefly consider a different class of conifold singularities
which can be regarded \ragana\ as the mirror geometries of the orbifolded
conifold spaces eq.\conifoldc\ considered before. Specifically, these 
six-dimensional generalized conifold singularities are defined as
\eqn\conifoldg
{{\cal G}_{KL}:\quad xy=u^Kv^L.}
As usual we easily determine the field theory from the underlying \cN=1 models
\ruranga.
The gauge group for $N$ `self-dual' D3 branes is given by
\eqn\gaugegen
{G=\bigotimes_{j=1}^{K+L}
\left({\rm SU(N)}_e\times {\rm SU(N)}_m\right).}
Without going into too many details, there are bifundamental massless fermions
and scalars in the $(\funda,\bfunda)$ of adjacent gauge group factors. 
Furthermore, depending on different choices of B-field backgrounds, there are
adjoint scalar multiplets. Finally, there are as usual
massless fermions in electric/magenetic bifundamentals of gauge groups with
identical indices.

\newsec{The supergravity description and T-duality}

\subsec{D3-branes at transversal singularities}

In this section we want to discuss the general form of the supergravity
solutions
for D3-branes at transversal singularities in type 0 string constructions.
Following the discussion in \roz\
we will assume that the background is asymptotically
of the form $M_4\times Y_6$,
where $M_4$ is flat Minkowski space-time, corresponding to the D3
world volume, and $Y_6$ is the non-compact transversal space with a 
singularity
at some fixed locus. 
In case of the six-dimensional orbifold and conifold singularities, 
we considered
in the last section, $Y_6$ is given as a cone over a five-dimensional
compact manifold $X_5$, the so-called horizon. Locally the metric on $Y_6$
can then be written in the  form
\eqn\tduala{dy_6^2=dr^2+r^2~d\Omega_X^2,  }
where $r$ is the radial coordinate on $Y_6$.
Asymptotically, for large and small $r$ the combined, 
near-horizon metric of the
D3-branes at $Y_6$ should take the form $AdS_5\times X_5$. In between,
the near-horizon metric is given by a warped product $M_4\times \tilde Y_6$,
where $\tilde Y_6$ is a fibration of $X_5$ over the radial coordinate $r$,
in general being different from the original transversal space $Y_6$.

Explicitly we consider the following ansatz for the 
10-dimensional metric metric \refs{\rkletsya,\rkletsyb,\roz}:
\eqn\tdualb{\eqalign{
ds^2 &= e^{{1\over 2}\phi(\rho)}ds_E^2 \cr
ds_E^2&=  {\rho^{-{1\over2}}\over
\sqrt{N} } e^{-{1\over 2}\xi(\rho)}
dx^\mu d^\mu
+\frac{\sqrt{N}}{16\rho^2}
e^{\frac{1}{2}\xi(\rho)-5\eta(\rho)}
d\rho^2+\sqrt{N}e^{\frac{1}{2}\xi(\rho)-\eta(\rho)}d\Omega_X^2,\cr }}
where $\xi(\rho)$ and $\eta(\rho)$ are functions which depend on the radial
coordinate $\rho$, and $\phi(\rho)$ is the dilaton.
These functions, together with four-form gauge fields have to be determined
as solutions of the equation of motions of the  type 0 string effective
action.
They depend crucially on the potential of the tachyon field which
is present in the type 0 string models.
This tachyon potential contains the following typical terms:
\eqn\tdualc{
h(T)=m^2f(T)+\frac{n^2}{f(T)},\quad f(T)=1+T+\frac{T^2}{2}.}
Here, $m$ is the number of electric D3-branes and $n$ the number
of magnetic D3-branes.
Like before we consider the two cases of either purely electric (or purely
magnetic) D3-branes or self-dual D3-branes.

\vskip0.2cm
\noindent{\it (i) Self-dual D3-branes}
\vskip0.2cm

In this case we have $m=n=N$.
The corresponding tachyon equations of motion are solved for vanishing
tachyon field $T=0$.
Hence the tachyon drops out from all the remaining equations of motion which 
are then identical to the equations of motion of $N$ D3-branes at a 
transversal singularity $Y_6$ in the corresponding type IIB superstring.
Namely, for self-dual D3-branes the near horizon metric is precisely of the
form $AdS_5\times X_5$, which corresponds to vanishing
functions $\xi$ and $\eta$ and constant dilaton field.
After a change of variables $\rho=r^{-4}$, the metric can be written as
\eqn\tduald{
ds_E^2=\frac{r^2}{\sqrt{N}}dx^\mu dx^\mu
+\frac{\sqrt{N}}{r^2}(dr^2+r^2d\Omega_X^2) }
This is precisely the near horizon metric of $N$ self-dual D3-branes at
the six-dimensional, transversal singularity $Y_6$, i.e. the full D3-brane
metric is obtained by replacing $\frac{N}{r^4}$ by the harmonic function
$H=1+\frac{N}{r^4}$ (suppressing the factors $g_s$ and $\alpha'$).

Let us study the orbifold singularities in a little more detail.
The horizon $X_5$ is just the five-sphere modded
out by the relevant discrete group $\Gamma$: $X_5=S^5/\Gamma$. 
Defining $S_5$ via its embedding in $R^6$ by the equation
$|z_1|^2+|z_2|^2+|z_3|^2=1$, the group $\Gamma$ acts on the three complex
coordinates $z_i$ precisely as discussed in section 2.
Consider e.g. the simplest
\cN=2 supersymmetric orbifold singularities where 
$\Gamma=\ZZ_K$ acts as $z_1\rightarrow e^{2\pi i/K}z_1$,
$z_2\rightarrow e^{-2\pi i/K}z_2$, $z_3\rightarrow z_3$.
Replacing the singularity at $z_1=z_2=0$ by a smooth 2-sphere,
$Y_6$ is given by the product $\IC\times ALE_{K-1}$, where the 
four-dimensional spaces $ALE_{K-1}$ describe the resolutions of the
$A_{K-1}$ singularities  and  
are complex two-dimensional non-compact relatives of $K3$, i.e. 
non-compact Ricci-flat hyper-K\"ahler manifolds.\footnote{$^3$}{All 
other singularities, like the
\cN=1 supersymmetric orbifolds or the conifolds can be constructed
as fibrations of (two) ALE spaces.} 
The ALE manifolds of the 
$A_{K-1}$ series correspond to the metrics given by the Gibbons-Hawking 
multi-center ansatz
\eqn\tduale{
dy_4^2=V(\vec{x})d\vec{x}^2+
V^{-1}(\vec{x})(d\tau+\vec{\omega}\cdot d\vec{x})^2 }
with the self-duality condition 
$\vec{\nabla}V=\vec{\nabla}\times \vec{\omega}$, and
\eqn\tdualf{
V=\sum_{i=1}^{K}\frac{1}{|\vec{x}-\vec{x_i}|}. }
This space $M_{K-1}$ is the smooth resolution of the singular variety
$xy=z^K$ in $\IC^3$ of type $A_{K-1}$ 
with $\partial M_{K-1}=S^3/\ZZ_K$.
The singular situation corresponds to the pol-terms coalescing:
$V=\frac{K}{|\vec{x}|}$.

On the other hand, before we have constructed the transversal singularity 
as a simple $\ZZ_K$ orbifold. Hence, 
we like to briefly recall (for the case $K=2$) that in fact the ALE-metric 
\tduale\
and the metric of the orbifold $\IC^2/\ZZ_K$ can be obtained from each other by 
a simple coordinate transformation.
For $K=2$, the two-center Gibbons-Hawking metric \tduale\ takes
the following explicit form
\eqn\tdualg{\eqalign{
dy_4^2&=\Biggl( \frac{1}{R_+}+\frac{1}{R_-}\Biggr)^{-1}\lbrack\tau+
\Biggl(\frac{z_+}{R_+}+\frac{z_-}{R_-}\Biggr)d~\tan^{-1}(y/x)\rbrack^2 \cr
&+\Biggl(\frac{1}{r_+}+\frac{1}{r_-}\Biggr)\lbrack dx^2+dy^2+
dz^2\rbrack ,\cr}}
where $z_\pm=z\pm z_0$, $R_\pm^2=x^2+y^2+z_\pm^2$.
The necessary coordinate transformations \rprasad\ are now given by
\eqn\tdualh{\eqalign{
x&=\frac{1}{8}\sqrt{r^4-64z_0^2}\sin\theta\cos\psi,\cr
y&=\frac{1}{8}\sqrt{r^4-64z_0^2}\sin\theta\sin\psi,\cr
z&=\frac{r^2}{8}\cos\theta,\cr
\tau&=2\phi . \cr}}
Using these new coordinates the metric \tdualg\ transform into
the following expression:
\eqn\tduali{\eqalign{
dy_4^2&=\frac{1}{4}r^2\lbrack 1-\frac{64z_0^2}{r^4}\rbrack\lbrack d\psi+\cos
\theta d\phi\rbrack^2 \cr
&+\lbrack 1-\frac{64z_0^2}{r^4}\rbrack^{-1}dr^2+\frac{1}{4}r^2\lbrack 
d\theta^2+\sin^2\theta d\phi^2\rbrack .\cr }}
This is precisely the Eguchi-Hanson metric. 
In the singular situation, $z_0=0$, this metric just describes the orbifold
space $\IC^2/\ZZ_2$, since the range of $\phi$ is $0\leq\phi <\pi$.
Note that resolving the singularity, i.e.
$z_0\neq 0$, the metric of the transversal space $Y_6$ is
not anymore of the form \tduala, 
but there are corrections of the form
\eqn\tdualj{
dy_6^2=r^2\biggl(\frac{dr^2}{r^2}+d\Omega_X^2+O(\frac{z_0^2}{r^2})\biggr).}

\vskip0.2cm

\vskip0.2cm
\noindent {\it (ii) Electric D3-branes}

\vskip0.2cm

Next we wish to discuss briefly  the supergravity solutions for entirely
electric D3-branes, i.e. we have $m=N$, $n=0$ in the tachyon potential.
Now the tachyon equations of motion are solved for non
vanishing tachyon field: $T=-1+\dots$. As a consequence, the ten-dimensional
metric is only asymptotically in the UV of the form $AdS_5\times S^5$;
for general values of $\rho$ 
the functions $\xi(\rho)$ and $\eta(\rho)$ are non-trivial, and the 
dilaton now
becomes a radius dependent function.
However the solution cannot be given anymore in closed form, but they can be 
expanded  around the 
asymptotic UV ($\rho\rightarrow 0$) $AdS_5\times 
S^5$ solution.
Explicitly, the first terms in this expansion are given as
\eqn\tdualj{
        \xi=-(\log\rho)^{-1},\quad \eta=-(\log\rho)^{-1},\quad
        \phi=\log(\frac{2^{13}}{27N})-2\log(-\log\rho).}
Due to the presence of these non-trivial functions, the ten-dimensional
metric looks like the metric of D3-branes now at a transversal
six-dimensional space $\tilde Y_6$, 
which only asymptotically in the UV is given by the original singularity
$Y_6$, but in general has the form of being
a warped product
over the five-dimensional horizon $X_5$ ($\rho=r^{-4}$):
\eqn\tdualk{\eqalign{
ds_E^2&=e^{-\frac{1}{2}\xi(r)}\frac{r^2}{\sqrt{N}}dx^\mu dx^\mu+
e^{\frac{1}{2}\xi(r)}\frac{\sqrt{N}}{r^2}d\tilde y_6^2, \cr
d\tilde y_6^2=&e^{-5\eta(r)}dr^2+e^{-\eta(r)}d\Omega_X^2.}}
For example, the `warped' Eguchi-Hanson metric, which asymptotically is 
identical to the resolution of the $\ZZ_2$ orbifold singularity,
takes the following form:
\eqn\tduall{\eqalign{
    d\tilde y_4^2 &=\frac{1}{4}r^2e^{-\eta(r)}
       \lbrack 1-\frac{64z_0^2}{r^4}\rbrack\lbrack d\psi+\cos
       \theta d\phi\rbrack^2 \cr
     &+e^{-5\eta(r)} \lbrack 1-\frac{64z_0^2}{r^4}\rbrack^{-1}dr^2+
    \frac{1}{4}r^2e^{-\eta(r)}\lbrack d\theta^2+
      \sin^2\theta d\phi^2\rbrack .\cr}} 
Using the change of coordinates in \tdualh\
it is straightforward to transform this metric into a  metric which
asymptotically approaches the 
two-center Gibbons-Hawking metric in \tdualg.

\subsec{T-dual supergravity description -- D4-branes and NS 5-branes}

In the last section we have described the supergravity picture
of $N$ D3-branes sitting on a six-dimensional transversal singularity
$Y_6$ (resp. $\tilde Y_6$) in \typeo\ string theory.
Now we like to discuss the T-dual picture of D4-branes which are intersected
by a certain number of NS 5-branes.
In order to keep the discussion simple we only consider the 
supergravity metric which
is T-dual to  D3-branes at the $\ZZ_K$ orbifold singularity.
As it is known, in this case the $A_{K-1}$ ALE spaces are T-dual to $K$ 
parallel NS 5-branes \rbersh . 

\vskip0.2cm
\noindent{\it (i) Self-dual D3-branes}

\vskip0.2cm

As discussed before, in the case of self-dual D3-branes the solutions of
the \typeo\ supergravity field equations are the same as in the corresponding
type IIB theories, namely they are given by the $AdS_5\times X_5$
background geometries.
Therefore, also the T-duality transformation acts precisely
in the same way as in the
\typeb\ superstring; the $N$ self-dual D3-branes will become after
the T-duality to type 0A $N$ electric plus $N$ magnetic D4-branes,
and the $\ZZ_K$ orbifolds are T-dualized into $K$ parallel NS 5-branes.
The T-duality is most conveniently studied using the 
Gibbons-Hawking ansatz \tduale\
for the metric of the transversal space.
Explicitly, 
the T-duality with respect to the $U(1)$-isometry generated by the Killing
vector $\partial /\partial \tau$ ($x_4$-direction)
gives with the well-known 
Buscher formula
the conformal flat metric of $K$ parallel, extremal NS 5-branes 
\eqn\tdualm{\eqalign{
           ds^2&=V(\vec{x})(d\tau^2+d\vec{x}^2),\cr
           B_{0i}&=\omega _i,\cr
           e^{2\phi}&=V(\vec{x}), }}
where the self-duality condition for the original metric is now, in the 
new axion-dilaton sector, assuring the condition for an axionic instanton
\eqn\tdualn{
H_{\mu\nu\rho}=
\sqrt{g}{\epsilon _{\mu\nu\rho}}^{\sigma}\partial _{\sigma}\phi .}
It is clear that the singular orbifold limit $\vec x_i\rightarrow 0$
corresponds to the situation where all NS 5-branes are
stacked
on top of each other.
Finally, putting together the $N$ D4-branes 
(in the 01234-directions) intersected by the $K$ NS 5-branes
(in the 012389-directions), the complete ten-dimensional
metric is given by the following expression:
\eqn\tdualp{
ds^2\sim\frac{r^2}{\sqrt{N}}\lbrack dx^\mu dx^\mu+V(\vec x)dx_4^2\rbrack
+\frac{\sqrt{N}}{r^2}\lbrack V(\vec x)dx_{5,6,7}^2+dx_{8,9}^2\rbrack .}

\vskip0.2cm
\noindent {\it (ii) Electric D3-branes}

\vskip0.2cm

The background spaces are now given by the deformed geometries
which only approach in the asymptotic UV region $AdS_5\times X_5$.
Nevertheless the deformed transversal spaces $\tilde Y_6$ still possess the 
same $U(1)$ isometry direction as before which can be used to perform the
T-duality transformation. Therefore, in the T-dual picture 
the electric D3-branes will be transformed into electric D4-branes, and
the deformed transversal singularities $\tilde Y_6$
will T-dualize into deformed five-dimensional NS branes, which only in
the UV region are described by the standard NS 5-brane metric.
To be explicit consider again the $\ZZ_K$ orbifold singularity.
Since its deformation by the functions $e^{-\xi}$ and $e^{-5\eta}$ 
can be best studied using polar coordinates,
the T-duality transformation will be most suitably now
also performed within this parametrisation.
For example, T-dualizing the 
deformed Eguchi-Hanson metric \tduall\ in the string frame
with respect to the isometry direction $\phi$, 
the metric for two deformed, parallel NS 5-branes can be explicitly 
constructed.

\newsec{Type 0 Hanany-Witten constructions}

In the last chapter on explicit supergravity solutions we have argued
that as in the type IIB superstring one can T-dualize
the type 0B D3-branes at transversal singularities into type 0A 
D4-branes intersected by a web of NS 5-branes. 
In case of dyonic  D4-branes the solutions are in fact identical
to the corresponding type IIA solutions, whereas for purely electric D4-branes
the type IIA solutions are only valid in the asymptotic UV region.

Using these results we now set up the type 0 Hanany-Witten \rhanwit\ rules
to obtain non-supersymmetric gauge theories from
D4-branes positioned into a web of NS 5-branes.
Dealing with dyonic D4-branes, the  non-supersymmetric Hanany-Witten 
rules we derive 
seem to be completely reliable, since we are confident that  the
corresponding backgrounds are indeed stable solutions of the
supergravity field equations.
In this case we can even hope to derive quantum corrections
in gauge theories, related to
the bending of the NS 5-branes, via the embedding of the brane configurations
into the type 0 version of M-theory.
In fact, many 
perturbative properties like the $\beta$-function coefficients
will not be changed comparing type II constructions with dyonic type 0
constructions, as we have already seen in the previous sections. 
Moreover it is quite conceivable that also non-perturbative duality 
symmetries,
like Seiberg-duality or S-duality, can be carried over from 
the underlying supersymmetric type II gauge theories to the non-supersymmetric
type 0 gauge theories \rschmaltz. 
For example, Seiberg duality can be again viewed as 
a particular motion of the NS 5-branes. 

The result that self-dual D3-branes can end on NS 5-branes is also 
supported by the recently discussed dualities of type 0A/B \rbergman. There
is was conjectured that the strong coupling limit of type 0A is
M-theory on $S^1/(-1)^{F_s} S$ where $S$ denotes the half-shift along the 
circle. In \rbergman\ is was argued that this  implies that the strong 
coupling  limit of type 0B on an $S^1$
is M-theory on $T^2/(-1)^{F_s} S$. Analogous to \typeb\ this suggests 
an $SL(2,\ZZ)$ self-duality of the \typeo, under which the fundamental
string is exchanged with the self-dual D1-brane. By definition a
fundamental string can end on a dyonic D5-brane. Applying S-duality
leads to a dyonic D1-brane ending on a NS 5-brane. By T-duality along
internal directions of the NS 5-brane one arrives at a self-dual
D3-brane ending on a NS 5-brane.  

Let us consider  the simplest NS 5-brane configuration, namely $K$ NS 5-branes
in the (012389)-directions, with suspended D4-branes which stretch along the
(01234)-directions. Hence this set up is precisely like the type IIA, \cN=2
Hanany-Witten configurations, discussed in \rwitten.
In case of a compact $x_4$ direction (elliptic models) the NS 5-branes are 
T-dual to the \cN=2, $\ZZ_K$ orbifold singularities. More precisely, the
T-dual of N D3-branes at the $\ZZ_K$ orbifold singularities leads
to an elliptic Hanany-Witten construction with an equal number of N 
D4-branes suspended between each pair of NS 5-branes. However, we wish to
generalise this picture by allowing also for a different number of D4-branes
in each 5-brane interval which corresponds
after T-duality  to having fractional branes, being partly wrapped around 
non-trivial cycles of the geometric singularity. In addition we will also
in general allow for non-compact $x_4$ directions. 

The massless spectrum simply follows from the T-duality to the
self-dual D3-branes on the $\ZZ_K$ orbifold singularity.
Alternatively, one can also directly start with the
\cN=2 supersymmetric type IIA Hanany-Witten construction and apply
the space-time fermion number projection $(-1)^{F_s}$
on the massless \cN=2 spectrum.
At the level of the effective gauge theories this projection should
be equivalent to the modding by the $\ZZ_2$ R-symmetry, which is discussed in
\rnekr.
This projection will not alter the one-loop $\beta$-function of the gauge 
theories.

Then, consider $K$ parallel NS 5-branes, and $n_j$
D4-branes, being suspended  in the $x^4$
direction between the $j^{\rm th}$
and $j^{\rm th}+1$ NS 5-brane. 
In case of non-elliptic models, 
for $j=0$ or $j=K$ the D4-branes are semi-infinite to the left 
of the first NS 5-brane or,
respectively, to the right of the $K^{\rm th}$ 
NS 5-brane.
The 4-branes are located in $x^8$-$x^9$ plane at the complex coordinate 
$v=x^8+ix^9$.
The  four-dimensional gauge group is  given by
\eqn\thwa{
  G=\bigotimes_{j=1}^{K-1} {\rm SU}(n_j)\times 
   \bigotimes_{j=1}^{K-1} {\rm SU}(n_j) .}
(For the elliptic models there is one more pair of gauge groups.)
The gauge coupling constants $g_j$ of every SU$(n_j)$ group factor
are determined by the differences between the positions of the NS 5-branes:
\eqn\hwc{
\frac{1}{g^2_j}=\frac{x^4_{j+1}-x^4_{j}}{g_s}, }
where $g_s$ is the string coupling constant.
Applying the fermion number projection on the corresponding \cN=2, type IIA
spectrum, we obtain massless states in the following representations
\eqn\hana{\eqalign{ 
         &\bigoplus_{j=1}^{K-1}  
        \left\{ \zwei{{\bf\rm Adj}_j}{1} + \zwei{1}{{\bf\rm Adj}_j} \right\}_B
     + \bigoplus_{j=1}^{K-1} 
        \left\{ 2\zwei{\funda_j}{\bfunda_j} + 
                2\zwei{\bfunda_j}{\funda_j} \right\}_F + \cr
     &\bigoplus_{j=1}^{K-2} \left\{ 2\vier{\funda_j}{\bfunda_{j+1}}{1}{1}+
                 2\vier{1}{1}{\funda_j}{\bfunda_{j+1}}
                 \right\}_B +\cr
     &\bigoplus_{j=1}^{K-2} \left\{
                \vier{\funda_j}{1}{1}{\bfunda_{j+1}}+
                \vier{\bfunda_j}{1}{1}{\funda_{j+1}}+
                \vier{1}{\funda_{j+1}}{\bfunda_{j}}{1}+
                \vier{1}{\bfunda_{j+1}}{\funda_{j}}{1}
                 \right\}_F \cr }}
The one-loop $\beta$-function coefficient is given by
\eqn\hwb{
b_1=2n_j-n_{j-1}-n_{j+1}. }
This is precisely the same number as in the \cN=2 parent models.
Note, that in the spirit of \rwitten\ this is consistent with the fact that,
since the supergravity
theory for self-dual D3-branes is the same as in \typeb, the bending
of the NS 5-branes in \typeo\ is the same, too.  Thus, also in the \typeo\
Hanany-Witten models one can derive the logarithmic running of the
gauge coupling constants from the bending of the NS 5-branes.  
Since the 5-brane background preserves \cN=2 space-time supersymmetry, the
resulting \cN=0 spectrum is automatically anomaly free.
We would like to mention that in the case of $K=2$ the two-loop 
$\beta$-function coefficient of the  \cN=0 theory 
accidentally vanishes, as well. Since the matter representations for the
bosons and fermions 
are different for the \cN=2 and the related \cN=0 gauge theory of course one 
does not expect to have vanishing \cN=0 $\beta$-function coefficient for
higher loops, as well.\pano
Analogous to the Coloumb moduli space in the \cN=2 gauge theories,
the motion of the dyonic D4-branes in the (89) direction is
related to giving vacuum expectation values to the adjoint complex scalars.
Since the couplings in the \cN=0 theory are the truncation  of the couplings
in the \cN=2 theory one has the following commuting diagram:
\bigno
\centerline{
$
\diagram 
                 \matrix{ {\cal N}=2 \cr {\rm SU}(K) }
                 & \rTo^{\langle\phi\rangle\neq 0} 
               & \matrix{ {\cal N}=2 \cr {\rm U}(1)^{K-1}}
                 \\ 
                  \dTo^{(-1)^{F_s}} & & \dTo_{(-1)^{F_s}} \\
                 \matrix{ {\cal N}=0 \cr {\rm SU}(K)^2}
               &   \rTo_{\langle\phi\rangle\neq 0} & 
                 \matrix{ {\cal N}=0 \cr {\rm U}(1)^{2K-2}} \\
\enddiagram               
$}
\bigno
Note that there does not exist further scalars in \hana\ which could 
describe the 
breaking up of a dyonic D4-brane into a D4 and a D$4'$-brane. This 
is consistent with the fact that all arguments given for the existence
of \typeo\ Hanany-Witten models relied on having dyonic
D4-branes. 
In addition to the Coulomb branch there exists also a Higgs branch
which corresponds to giving vev's to the bifundamental scalar fields.
In this way a pair of electric or magnetic  SU(N)$^2_{e,m}$ gauge groups
is higgsed to its diagonal subgroup SU(N)$_{e,m}$. 
In the corresponding Hanany-Witten
brane picture this Higgs moduli space is related to the motions of
the NS 5-branes, whereas in the T-dual orbifold picture the bifundamental scalar
vev's correspond to the resolution parameters (2-spheres) of the
$\ZZ_K$ orbifold singularity. Since the electric and magnetic D-branes
feel these parameters, which are related to the orbifold respectively NS
5-brane background, in the same way, it is plausible that in the field theory
the vev's of the electric and magnetic scalar fields must be the same
such that the electric and magnetic gauge groups are higgsed simultaneously.

It is straightforward to generalise the Hanany-Witten set up of parallel
NS 5-branes to type 0 models which contain several NS 5-branes
with different kind of orientations. In particular brane box-models \rhanzaf\
with D5-branes in the (012348)
directions which fill boxes of NS 5-branes  in the
(012389) directions and of NS$'$ 5-branes in the (012345) directions
follow by applying two T-dualities with respect to $x_4$ and $x_8$
on the transversal $\ZZ_K\times \ZZ_L$ orbifold singularity. These
gauge theories  are in general chiral and potentially lead to dangerous
gauge anomalies.

The orbifolded conifold singularities in eq.\conifoldc\ are T-dual
(T-duality in $x_4$ and $x_8$) to diamonds of NS 5-branes \ragana.
By rotating the diamonds into certain directions, the $\ZZ_K\times \ZZ_L$
brane box models are rediscovered.
On the other hand, the mirror conifold geometries eq.\conifoldg\ lead
after a single T-duality in $x_6$ to rotated NS brane configurations 
\ruranga ,
namely to Hanany-Witten models, where electric and
magnetic D4-branes, along
the (01236) directions, are suspended
between intervals of
$K$ NS 5-branes in the original (012389) directions and $L$
rotated NS$'$ 5-branes along the (012367) directions. The specific ordering
possibilities of NS and NS$'$ branes correspond to different choices
of B-field background. For $K=L=1$, $N_c$ finite
D4 branes and $N_f$ semi-infinite D4-branes, one can easily construct a
(non-elliptic) non-supersymmetric
QCD type of model with gauge group $SU(N_c)_e\times SU(N_c)_m$ and 
$N_f$ massless, fundamental  matter fermions and  scalars
plus massless fermions 
in bifundamental gauge representations.\footnote{$^4$}{This non-supersymmetric
model and its Seiberg-dual was recently discussed in \rkol.}
As usual, the 1-loop $\beta$-function is inherited from the \cN=1
parent model: $b_1=3N_c-N_f$. 
As discussed in \relitzur , this class of brane models is particularly useful
to study the Seiberg-duality in the corresponding \cN=1, type IIA brane
picture by moving an appropriate number of NS
or NS$'$ branes through the D-branes.
Since the correlation functions in the large N limit of
the type 0 models agree with those
of their type II parents, it is very conceivable that carrying out the 
same movements of NS-branes in the non-supersymmetric type 0 models
corresponds to a non-perturbative large N Seiberg-duality 
in non-supersymmetric
field theories. Indeed, one can show along these lines that there exists a dual, 
non-supersymmetric
QCD model with $\tilde N_c=N_f-N_c$ and $\tilde N_f=N_f$.

\newsec{Summary}

This paper provides a systematic study of non-supersymmetric gauge theories
from electric/magnetic D-branes in type 0 constructions. Since these 
models share a lot of properties with their \cN=2,1 supersymmetric 
counterparts from type II superstrings, their 
non-perturbative dynamics most likely exhibits a lot of intriguing features,
like certain kinds of non-perturbative duality symmetries.
In addition, 
we expect that these models in spite of being non-supersymmetric
nevertheless possess very nice renormalization properties that could make
them appealing also from the phenomenological point of view. 
The ultimate hope would be to gain some information about the higher loop
and non-perturbative behaviour by embedding the 
non-supersymmetric Hanany-Witten configurations
into M-theory.
It would be also interesting to study D3-branes on 
non-supersymmetric singularities in greater detail and to find out
to which kind of non-supersymmetric Hanany-Witten configurations they
are mapped to.

\bigskip\bigskip\centerline{{\bf Acknowledgements}}\pano
We thank L. G\"orlich for discussion and
A.F. thanks CDCH-UCV for a research grant 03.173.98.
The  work is partially supported by the E.C. project
ERBFMRXCT960090.
During the final
stages of this work we became aware of \rkol, where Hanany-Witten and
brane box models in \typeo\ were discussed, as well.

\vfill\eject

\listrefs
\bye